%% file: paper.tex
\documentclass[]{ceurart}
\usepackage{listings}
\usepackage{bm}
\usepackage{algorithm, algpseudocode}
\algnewcommand{\LineComment}[1]{\State \(\triangleright\) #1}
\usepackage{wrapfig}
\usepackage{courier}
\usepackage{ragged2e}

\newtheorem{definition}{Definition}[section]
\newtheorem{example}{Example}[section]

\newcommand{\nop}[1]{{}}

\def\calc{\mathcal{C}}
\def\cala{\mathcal{A}}

\def\cali{\mathcal{I}}
\def\calt{\mathcal{T}}
\def\calv{\mathcal{V}}

\def\call{\mathcal{L}}

\def\call{\mathcal{L}}

\def\calf{\mathcal{F}}

\def\rel{\mathit{rel}}

\def\calp{\mathcal{P}}

\def\avar{\textsf{AVar}}

\def\T{{\mathbf{T}}}

\def\S{\textbf{S}}

\newcommand\DILP{$\partial$ILP}

\begin{document}
\copyrightyear{2023}
\copyrightclause{Copyright for this paper by its authors.
  Use permitted under Creative Commons License Attribution 4.0
  International (CC BY 4.0).}

\conference{In A. Martin, K. Hinkelmann, H.-G. Fill, A. Gerber, D. Lenat, R. Stolle, F. van Harmelen (Eds.), 
Proceedings of the AAAI 2023 Spring Symposium on Challenges Requiring the Combination of Machine Learning and Knowledge Engineering (AAAI-MAKE 2023), Hyatt Regency, San Francisco Airport, California, USA, March 27-29, 2023.}

\title{PyReason: Software for Open World Temporal Logic}

\author{Dyuman Aditya}[%
orcid=0000-0002-4889-3499,
email=,
url=,
]
\fnmark[1]
\address{Arizona State University,
  699 S Mill Ave, Tempe, AZ, 85281, USA}

\author{Kaustuv Mukherji}[%
orcid=0000-0001-8044-1110,
email=kmukherji@asu.edu,
url=https://search.asu.edu/profile/4179815,
]
\cormark[1]
\fnmark[1]

\author{Srikar Balasubramanian}[%
orcid=,
email=,
url=,
]

\author{Abhiraj Chaudhary}[%
orcid=,
email=,
url=,
]

\author{Paulo Shakarian}[%
orcid=,
email=pshak02@asu.edu,
url=https://labs.engineering.asu.edu/labv2/,
]
\cormark[1]

\cortext[1]{Corresponding author.}
\fntext[1]{These authors contributed equally.}

\begin{abstract}
  The growing popularity of neuro symbolic reasoning has led to the adoption of various forms of differentiable (i.e., fuzzy) first order logic.  We introduce PyReason, a software framework based on generalized annotated logic that both captures the current cohort of differentiable logics and temporal extensions to support inference over finite periods of time with capabilities for open world reasoning.  Further, PyReason is implemented to directly support reasoning over graphical structures (e.g., knowledge graphs, social networks, biological networks, etc.), produces fully explainable traces of inference, and includes various practical features such as type checking and a memory-efficient implementation.  This paper reviews various extensions of generalized annotated logic integrated into our implementation, our modern, efficient Python-based implementation that conducts exact yet scalable deductive inference, and a suite of experiments. PyReason is available at: \textbf{\url{github.com/lab-v2/pyreason}}.
\end{abstract}

\begin{keywords}
  Logic programming \sep
  Neuro Symbolic Reasoning \sep
  Generalized annotated logic \sep
  Temporal logic \sep
  First order logic \sep
  Open world reasoning \sep
  Graphical reasoning \sep
  AI Tools
\end{keywords}

\maketitle

\section{Introduction}
\label{sec:introduction}

Various neuro symbolic frameworks utilize an underlying logic to support capabilities such as fuzzy logic~\cite{ltn22}, parameterization~\cite{lnn2020}, and differentiable structures~\cite{deepMinIlp2018}.  Typically, implementations of such frameworks create custom software for deduction for the particular logic used, which limits modularity and extensibility.  Further, emerging neuro symbolic use cases including temporal logic over finite time periods~\cite{ma_stlnet_2020} and knowledge graph reasoning~\cite{sen_logical_2022} necessitate the need for a logical frmaework that encompasses a broad set of capabilities.  Fortunately, generalized annotated logic~\cite{ks92} with various extensions~\cite{ssTAI22,tplp13,mancalog13} capture many of these capabilities.  \textbf{In this paper we present a new software package called PyReason for performing deduction using generalized annotated logic that captures many of the desired capabilities seen in various neuro symbolic frameworks including fuzzy, open world, temporal, and graph-based reasoning.}  Specifically, PyReason includes a core capability to reason about first order (FOL) and propositional logic statements that can be annotated with either elements of a lattice structure or functions over that lattice.  Further, we have provided for additional practical syntactic and semantic extensions that allow for reasoning over knowledge graphs, temporal logic, reasoning about various network diffusion models, and predicate-constant type checking constraints. This implementation provides for a fast, memory optimized, implementation of the fixpoint operator used in the deductive process. By implementing the fixpoint operator directly (as opposed to a black box heuristic) the software enables full explainability of the result.  As such is the case, this framework captures not only classical logic, but a wide variety of other logic frameworks including fuzzy logic~\cite{hohle1978probabilistic, alsina1983some, vojt01}, weighted real valued logic used in logical neural networks~\cite{lnn2020}, van Emden’s logic~\cite{vanemdenlogic}, Fitting's bilattice logic~\cite{fitting1990bilattices}, various logic frameworks for reasoning over graphs or social networks \cite{mancalog13,tplp13,sdnops13} (as well as the various network diffusion models captured by those frameworks), and perhaps most importantly, logic frameworks where syntactic structure can be learned using differentiable inductive logic programming~\cite{deepMinIlp2018,nttIlp21} as well as other neuro symbolic frameworks~\cite{lnnInduction22,ssTAI22}.  The key advantages of our approach include the following:
\begin{enumerate}
    \item\textbf{Direct support for reasoning over knowledge graphs.} Knowledge graph structures are one of the most commonly-used representations of symbolic data.  While black box frameworks such as \cite{thomasNsr21} also permit for reasoning over graphical structures, they do not afford the explainability of our approach.
    \item\textbf{Support for annotations.}
    Classical logic implementations such as Prolog~\cite{colmerauer1982prolog} and Epilog~\cite{schubert2000episodic} inherently do not support annotations or annotation functions, hence lack direct support for capabilities such as fuzzy operators.  Further, our framework goes beyond support for fuzzy operators by enabling arbitrary functions that can be used over real values or intervals of reals.  This is a key advantage to reasoning about constructs learned with neuro symbolic approaches such as \cite{lnn2020,deepMinIlp2018,nttIlp21,lnnInduction22,ssTAI22}.
    \item\textbf{Temporal Extensions.}  While the framework of \cite{ks92} was shown to capture various temporal logics, extensions such as \cite{mancalog13} have provided for syntactic and semantic add-ons that explicitly represent time and allow for temporal reasoning over finite temporal sequences.  Following \cite{mancalog13}, we use a semantic structure that represents multiple time points, but we have implemented this in a compact manner to preserve memory.  Our solution allows for fuzzy versions of rules such as ``if $q(A)$ then $r(A)$ in $t$ time steps.''  Note that these capabilities are not present in nearly every current implementation of fuzzy logic.
    \item\textbf{Use of interpretations.}  We define interpretations as annotated function over predicates and time together. It allows us to capture facts which are true before $t=0$. While annotated logic~\cite{ks92} can subsume various temporal logics without additional constructs, we have enabled temporal reasoning through incorporating a temporal component in interpretations. By combining annotated predicates and the time variable, we believe our framework is more flexible and suitable for emerging neuro symbolic applications involving time - as such applications will inherently require both time and real-valued annotations. Additionally, it is to be noted that we do not make a closed world assumption i.e. anything that is not mentioned in the initial set of interpretations is $false$. Instead, we consider all other interpretations to be unknown at the beginning of time.
    \item\textbf{Graphical Knowledge Structures.}  We also implement \cite{tplp13} which provides graphical syntactic extensions to \cite{ks92}.  This is included in our implementation, notably adding extended syntactic operators for reasoning in such structures (e.g., an existential operator requiring the existence of $k$ items).  An example of such a rule would be a fuzzy version of ``if $q(A)$ and there exist $k$ number of $B$'s such that $b(A,B)$ then $r(A)$''.\footnote{Note that while this example is classical, PyReason supports fully annotated logic, allowing for arbitarily defined fuzzy operators (e.g., t-norms); See section~\ref{sec:logical_framework} and online supplement for technical details.}
    \item\textbf{Reduction to computational complexity due to grounding.}  Our software leverages both the inherent sparsity of the graphical structure along with a novel implementation of predicate-constant type checking constraints that significantly improves utility in a variety of application domains but also provide drastic reduction to complexity induced by the grounding problem.  We are not aware of any other framework for first-order logic that provides both such capabilities.
    \item\textbf{Ability to detect and resolve inconsistencies in reasoning.} As logical inferences are deduced through applications of the fixpoint operator over predefined logical rules, logical inconsistencies can not only be detected but also located exactly where in the inference process the inconsistency occurred. We resolve any such inconsistencies by leveraging uncertainty. In the software implementation, as soon as an inconsistency is detected we relax and fix the bounds to complete uncertainty. The ability to check and locate inconsistencies enhance the explainability feature. Neuro symbolic approaches like~\cite{lnn2020,ssTAI22} may also look to leverage inconsistency as part of loss during the training phase.
\end{enumerate}
In section~\ref{sec:logical_framework}, we outline the syntax and semantics of \cite{ks92} as well as our extensions. Our software implementation is described in section~\ref{sec:software_implementation} and is expanded upon in the online only supplement. In section~\ref{sec:experimental_results}, we provide experimental results of our framework to demonstrate reasoning capabilities in two different real-world domains. We have conducted experiments on a supply-chain~\cite{yan_theory_2015} ($10K$ constants), and a social media~\cite{takac2012data} ($1.6M$ constants) dataset.  For evaluation, we used various manually-curated logic programs specifying rules for the temporal evolution of the graph, completion of the graph, and other such practical use-cases (e.g., identifying potential supply chain disruptions) and examined how various aspects affect runtime and memory usage (e.g., number of constants, predicates, timesteps, inference steps, etc.).  The results show that both runtime and memory remain almost constant over large ranges, and then scale sub-linearly with increase in network size.

\subsubsection*{Online Resources}
\label{sec:online_resources}
Open source python library is available at: \textbf{\url{pypi.org/project/pyreason}}. \\
\noindent PyReason codebase can be found  at: \textbf{\url{github.com/lab-v2/pyreason}}. \\
\noindent Online only supplement is available at: \textbf{\url{github.com/lab-v2/pyreason/tree/main/lit}}

\section{Logical Framework}
\label{sec:logical_framework}

In this section, we provide an overview of the annotated logic framework with a high-level description of the logical constructs, knowledge graph structure, key optimizations, and operation of the fixpoint algorithm.\\

\begin{figure}
    \begin{center}
        \includegraphics[width=0.7\linewidth]{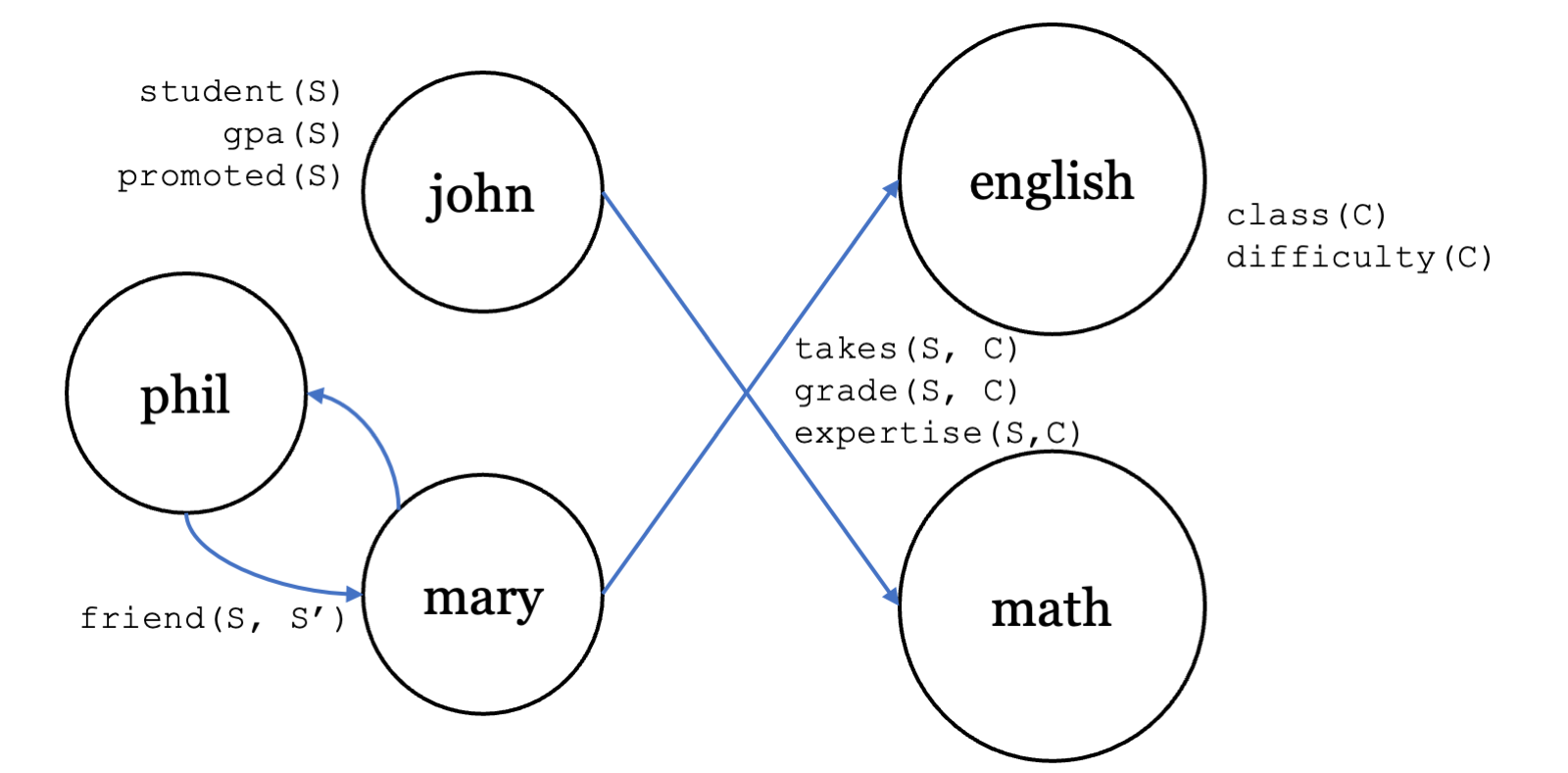}
    \end{center}
    \caption{\label{fig:example_block}Example of a knowledge graph}
\end{figure}

\noindent\textbf{Knowledge graph.}  We assume the existence of a graphical structure $G=(\calc,E)$ where the nodes are also constants (denoted set $\calc$) in a first-order logic framework.  The edges, denoted $E \subseteq \calc \times \calc$, specify whether any type of relationship can exist between two constants.  Similar to recent frameworks combining knowledge graphs and logic \cite{deepMinIlp2018, thomasNsr21}, we shall assume that all predicates in the language are either unary (which can be thought of as labeling nodes) or binary (which can be thought of as labeling edges).  We note that we assume the existence of a special binary predicate $\rel$, which we shall treat as a reserved word.  For $(a,b) \in E$ we shall treat $\rel(a,b)$ as a tautology and for $(a,b) \notin E$ we shall treat $\rel(a,b)$ as uncertain.  Note that we can support no restrictions among the pairing of constants by creating $G$ as a fully connected graph.  Likewise, we easily support the propositional case by using a graph of a single node (essentially treating unary predicates as ground atoms).  We provide a running example in this section.  In Figure~\ref{fig:example_block}, we illustrate how a knowledge graph is specified in our framework.\\
\begin{example}[Knowledge Graph]
Consider the following nodes: three students- Phil, John, Mary and two classes- English and Math. Nodes and edges have unary and binary predicates as shown in Fig.~\ref{fig:example_block}.
Hence we get the following non-ground atoms:

\texttt{student(S), 
gpa(S), 
promoted(S)}

\texttt{class(C), 
difficulty(C)}

\texttt{friend(S,S')}

\texttt{takes(S,C), 
grade(S,C), 
expertise(S,C)}

\noindent
Here, S, S', and C are variables which when grounded with constants from the graph, produce ground atoms such as:

\texttt{student(john), student(phil), student(mary)}

\texttt{class(math), class(english)}

\texttt{takes(john,math), takes(mary, english)}

\texttt{\ldots}

\noindent
In the propositional case, a non-ground atom reduces to a propositional statement. For e.g. The predicate ``takes(john,math)'' can be represented as a propositional statement: ``John takes Math class'' and can be either True or False. It is true in this example, as shown in Fig.~\ref{fig:example_block}.

\end{example}
\noindent\textbf{Real-valued Interval Annotations.}  A key advantage of annotated logic~\cite{ks92} is the ability to annotate the atoms in the framework with elements of a lattice structure as well as functions over that lattice.  In our software, we use a lower lattice structure consisting of intervals that are a subset of $[0,1]$.  This directly aligns with the truth interval for fuzzy operators~\cite{vojt01}, as well as paradigms in neuro symbolic reasoning~\cite{lnn2020, ssTAI22}, and social network analysis~\cite{tplp13, mancalog13}.  We can fully support scalar-valued annotations by simply limiting manipulations to the lower bound of the interval and keeping the upper bound set at $1$.  These annotations can support classical logic by limiting annotations to be $[0,0]$ (false) and $[1,1]$ (true).  It can also support tri-valued logic by permitting $[0,1]$, which represents no knowledge.  Of course, there is no need to conduct restrictions, especially if it is desirable to support logics that make full use of the interval \cite{lnn2020,tplp13, mancalog13}. Additionally, we support literals as detailed in~\cite{ssTAI22}. We treat negations the same way as in \cite{ltn22} - for an atom annotated with $[\ell,u]$, we annotate its strong negation($\neg$) with $[1-u,1-\ell]$.\\

\begin{example}[Real-valued Interval Annotations]
Continuing with the previous example, we can support a variety of annotations as described above. \\
\noindent Propositional logic:

\texttt{student(john): [1,1]} (example of a True statement)

\texttt{takes(mary,math): [0,0]} (example of a False statement)

\noindent Fuzzy logic (using scalar values):

\texttt{gpa(john): [X,1], X $\in$ [0,1]}

\noindent Full interval usage:

\texttt{difficulty(english): [0.3,0.7]} (both bounds are used here to capture the variation among students regarding the perceived difficulty of the subject ``english'').

\noindent Modeling uncertainty and/or tri-valued logic: \\
Let's assume that we do not have complete knowledge of this network - specifically, we do not have any information about the friendship between John and Phil. So, they might be friends (annotated [1,1]) or not friends (annotated [0,0]). Our framework can model such a case as:

\texttt{friend(john,phil): [0,1]}

\end{example}

\noindent\textbf{Interpretations.}  Commonly in logic frameworks, an initial set of facts is used. We use the term ``initial interpretations'' to capture annotations correct at the beginning of a program. In the envisioned domains - to include the ones in which we perform experiments - these initial interpretations shall be represented as a knowledge graph that not only includes graph $G$ but also attributes on the nodes and edges (resembling predicates) and real-valued interval annotations (specifying the initial annotations for each element). Additionally, following intuitions from various temporal logic frameworks that incorporate both temporal and other real-valued annotations~\cite{mancalog13,tplp13, dekhtyar99,soma07,apt11}, we extend our syntax to provide for temporal annotations as part of the interpretations. Following the related work, time is represented as finite discrete time-points. The initial interpretations comprises what is to be treated as true before time $0$.  Further, with the initial interpretations we can specify predicates as being either static (in other words, ground atoms formed with those predicates retain the same annotation across all time periods) or non-static (which are permitted to change).  The ability to add this restriction has clear benefit in certain domains, and also allows for key implementation efficiencies for reasoning across time periods. Further, it is noted various inductive logic programming paradigms~\cite{deepMinIlp2018, francca2014fast} utilize ``extensional'' predicates that are also unchanging - which could be treated as ``static'' in PyReason.

\noindent
\textbf{Syntax}:

$I(A,\hat{t}):[L,U]$ \\
where, $A$ can be an atom (propositional case) or predicate (first order logic), $\hat{t}$ is either the time point $T=t$ for which the interpretation $I$ is valid, or if the interpretation is static, i.e. remains unchanged for all time-points then $\hat{t}=s$. So,
\begin{equation}
    \hat{t}=
    \begin{cases}
      s, & \text{if}\ I(A,\hat{t})\ \text{is static} \\
      t, & t \in T\ \text{if}\ I(A,\hat{t})\ \text{is time-variant}
    \end{cases}
  \end{equation}
Annotation $[L,U] \rightarrow [0,1]$ (or, in propositional case $[L,U] \in {[0,0],[1,1]}$).
We incorporate literals in our system by having separate interpretations for an atom and its negation. We note that, excepting the case of static atoms,  ground atoms at different time points need not be dependent upon each other.  For example, atom ``a'' at time $1$ can be annotated with $[0.5,0.7]$ and annotated with $[0.1,0.2]$ at time 2.  There is no monotonicity requirement between time points.

\begin{example}[Interpretations]
Continuing the previous example, \\
\noindent Initial set of facts regarding student enrollment:

\texttt{I(student(john),0) = [1,1]} (John is enrolled as a student)

\texttt{I(student(mary),0) = [1,1]} (Mary is enrolled as a student)

\texttt{I(student(phil),0) = [0,0]} (Phil is not enrolled as a student)

\noindent Static interpretations can be used for always true facts like:

\texttt{I(class(english), s) = [1,1]} (English is a class offered at all time-points)

\noindent Using temporal annotation to capture variation over time:

\texttt{I(takes(john,math),1) = [1,1]} (John takes Math class at time $t=1$)

\texttt{I(takes(john,math),5) = [0,0]} (But is no longer taking Math at $t=5$)

\noindent All other interpretations, if unspecified at $t = 0$, are initialized with $[0,1]$.
\end{example}

\noindent\textbf{Logical Rules.}  Rules are the key syntactic construct that enables changes to atoms formed with non-static predicates. Historically logical rules had mostly been written by domain experts, until early work like Apriori~\cite{agrawal1996fast} and FOIL~\cite{quinlan1990learning} to learn association rules from data followed by the emergence of rule mining techniques like causal rule mining~\cite{stanton15} and annotated probabilistic temporal logic~\cite{soma07, shakarian2011annotated, shakarian2012annotated}. 
More recently, there has been research on Differentiable Inductive Logic Programming (\DILP) - an inductive rule learning method to learn logical rules from examples~\cite{deepMinIlp2018, nttIlp21, sen2022neuro}. 
In the below list $UnaSet$ and $BinSet$ are arbitrarily sets of unary and binary predicates relevant to the rules while $pred$ is always a non-static predicate.  Note that the total number of atoms in the body is assumed to be $n$ (across all different conjunctions).  The symbol $\exists_k$ means there exists at least $k$ number of constants such that the ensuing logical sentence is satisfied.

\begin{enumerate}
    \item Ground rule for reasoning within a single constant or edge:\\
\begin{small}
$pred(c):f(x_1,\ldots,x_n)\leftarrow_{\Delta t} \bigwedge_{pred_i \in UnaSet} pred_i(c):x_i$ \\

$pred(c,c'):f(x_1,\ldots,x_n)\leftarrow_{\Delta t} \bigwedge_{pred_i \in BinSet} pred_i(c,c'):x_i$ \\
\end{small}

    \item Universally quantified non-ground rule for reasoning within a single constant or edge:\\
\begin{small}
$\forall X: pred(X):f(x_1,\ldots,x_n)\leftarrow_{\Delta t} \bigwedge_{pred_i \in UnaSet} pred_i(X):x_i$ \\

$\forall X,X' \ s.t. \ (X,X')\in E: pred(X,X'):f(x_1,\ldots,x_n)\leftarrow_{\Delta t} \bigwedge_{pred_q \in BinSet} pred_q(X,X'):x_q\wedge \bigwedge_{pred_r \in UnaSet}pred_r(X):x_r \wedge\bigwedge_{pred_s \in UnaSet'}pred_s(X'):x_s $ \\
\end{small}

    \item Universally quantified non-ground rule for reasoning across an edge:\\
\begin{small}
$\forall X: pred(X):f(x_1,\ldots,x_n)\leftarrow_{\Delta t} \exists_k X': rel(X,X'):[1,1] \wedge  \bigwedge_{pred_q \in BinSet} pred_q(X,X'):x_q\wedge \bigwedge_{pred_r \in UnaSet}pred_r(X):x_r \wedge\bigwedge_{pred_s \in UnaSet'}pred_s(X'):x_s $\\
\end{small}

    \item Non-ground rule with rule based quantifier in the head:\\
\begin{small}
    $pred(X):[A_s(l_1,l_2,\ldots,l_n), A_s(u_1,u_2,\ldots,u_n)] \leftarrow \bigwedge_{X_i\ \text{s.t.}\ (X,X_i) \in E} pred'(X,X_i):[l_i,u_i]$
    
    Here, $A_{s,m}^k(S)$ could be the $m^{th}$ rule based quantifier defined over set $S$ such that,
    $A_{s,m}^k(S) = k^{th}\ \text{highest value in set}\ S$. \\
\end{small}
\end{enumerate}

\begin{example}[Logical Rules] For the continuing example we can formulate some interesting rules based on the formats given above as: \\

\begin{enumerate}
    \item $promoted(X):[T(l_1,l_2),U(u_1,u_2)] \leftarrow_{\Delta t=1} student(X):[l_1,u_1] \wedge gpa(X):[l_2,u_2]$

    which says, ``If $X$ is a student with bounds $[l_1,u_1]$ and has a gpa with bounds $[l_2,u_2]$, then $X$ is likely to be promoted, at the next timestep, with bounds given by a function of $[l_1,u_1]$ and $[l_2,u_2]$.''\\
    
    Here, $T$ could be a T-norm. Some well known examples of T-norms are:
    \begin{enumerate}
        \item Minimum: $T(a,b) = T_{min}(a,b) = min(a,b)$
        \item Product: $T(a,b) = T_{prod}(a,b) = a\cdot b$
        \item Łukasiewicz: $T(a,b) = T_{luk}(a,b) = max(0, a+b-1)$
    \end{enumerate}

    PyReason also supports other well known logical functions like $T-conorm$, algebraic functions like $max$, $min$, $average$, among others.\\
    
    \item $\forall X,Y \ expertise(X, Y):[0.6*L,1] \leftarrow_{\Delta t=0} grade[X,Y]:[L,1] \wedge student(X):[1,1] \wedge class(Y):[1,1]$

    which says, ``If $X$ is a student who obtains a grade $[L,1]$ in class $Y$, then we can estimate $X$'s expertise of subject $Y$ by defining an annotation function $[0.6*L,1]$ over a single annotation $[L,1]$.''\\
    
    \item $gpa(john):[\frac{x_1+x_2}{2},1] \leftarrow_{\Delta t = 0} \exists_{i=2} C_i \in \calc : class(C_i):[1,1] \wedge takes(john, C_i):[1,1] \wedge grade(john, C_i):[x_i, 1] $
    
    which says, ``If $john$ takes and earns grades for two classes, then his $gpa$ can be calculated using the algebraic function $avg$ in the head of the given existentially quantified ground rule.''\\
    
    \item $friend(S, S'):[1,1] \leftarrow_{\Delta t=2} takes(S,C):[1,1] \wedge takes(S',C):[1,1] \wedge class(C):[1,1]$
    
    a propositional rule with temporal extension which states, ``If two students $S$ and $S'$ take the same class $C$, they develop a friendship after two timesteps.''\\
    
    \item $\forall S, S', S'' \ friend(S, S''):[1,1] \leftarrow_{\Delta t=1} friend(S, S'):[1,1] \wedge friend(S', S''):[1,1]$
    
    an universally quantified non-ground rule analogous to the associative rule in mathematics which encapsulates, ``Having a common friend $S'$ leads to friendship between two people $S$ and $S''$.''
\end{enumerate}

\end{example}

\noindent\textbf{Fixpoint Operator for Deduction.} Central to the deductive process is a fixpoint operator (denoted by $\Gamma$) which has previously been proven to produce all atoms entailed by a logic program (rules and facts) in \cite{ks92, ssTAI22} and these results were extended for the temporal semantics in \cite{mancalog13,tplp13}.  It is noteworthy that this is an exact computation of the fixpoint, and hence providing the minimal model associated with the logic program allowing one to easily check for entailment of arbitrary formulae.  Further, the result is fully explainable as well: for any entailment query we would have the series of inference steps that lead to the result.  This differs significantly from other frameworks that do not provide an explanation for deductive results~\cite{thomasNsr21} though a key difference is that the reasoning framework implemented in PyReason allows for exact and efficient polynomial time inference, while others have an intractable inference process.

\begin{example}[Fixpoint Operator($\Gamma$)]
Consider we have the following set of initial interpretations in addition to the ones specified before:
\begin{flushleft}\texttt{I(takes(john,english),1) = I(takes(john,english),2) = [1,1]}

\texttt{I(takes(mary,english),2) = I(takes(mary,english),3) = [1,1]}

(John takes English at t=1,2 and Mary takes English at t=2,3)

\texttt{I(friend(mary,phil),s) = [1,1]}

(Mary and Phil are friends for the entire time considered)
\end{flushleft}
\begin{flushleft}
\noindent And we consider the rule set $\bm{R}$ to be made of rule 4 and 5 from above. We initialize:

\texttt{$\forall$S,S' I(friend(S,S'),0) = [0,1]}
(all $friend$ relationships initialized as unknown)

\noindent and then update:

\texttt{I(friend(mary,phil),s) = [1,1]}
(from initial interpretations)
\end{flushleft}
\begin{flushleft}
\noindent Application of\, $\Gamma$ at T=0 and 1 yields no change in $\bm{I}$ as none of the rules are fired.

\noindent At T=2, rule 4 fires with the following groundings:

$friend(john, mary):[1,1] \leftarrow_{\Delta t=2} takes(john,english):[1,1] \wedge takes(mary,english):[1,1] \wedge class(english):[1,1]$

$friend(mary, john):[1,1] \leftarrow_{\Delta t=2} takes(mary,english):[1,1] \wedge takes(john,english):[1,1] \wedge class(english):[1,1]$

\noindent This would result in a change in $\bm{I}$ at T = 4, as $\Delta t=2$ for the rule above and it is fired at T=2.

\texttt{I(friend(john,mary),4) = [1,1]}

\texttt{I(friend(mary,john),4) = [1,1]}

\noindent At T=3, as $\bm{I}$ is still unchanged, application of\, $\Gamma$ does not lead to any of the rules firing.

\noindent At T=4, application of\, $\Gamma$ with the updated interpretation leads to firing of grounded rule 5 as:

$friend(john, phil):[1,1] \leftarrow_{\Delta t=1} friend(john, mary):[1,1] \wedge friend(mary, phil):[1,1]$

\noindent And results in:

\texttt{I(friend(john, phil),5) = [1,1]}
\end{flushleft}

\noindent The above illustrates how PyReason makes logical inferences by exact application of the fixpoint operator($\Gamma$). In this example, we are able to trace how the interpretation \texttt{I(friend(john, phil),t)} changed over time, and which rules caused these changes. This shows that this process is completely explainable, and can be leveraged in emerging neuro symbolic applications.
\end{example}

\noindent\textbf{Constant-Predicate Type Checking Constraints.}  Key to reducing the complexity and speeding up of the inference process is type-checking. We leverage the sparsity commonly prevalent in knowledge graphs to significantly cut down on the search space during the grounding process. We noticed that typically a graph will have nodes of different types, and predicates typically were defined only over constants of a specific type. While initializing the interpretations, type checking takes this into account and only creates ground atoms for the subset of predicate-constant pairs which are compatible with each other. However, we note that this is an option, as in some applications such information may not be available.

\begin{example}[Constant-Predicate Type Checking] In the continuing example we see that the predicates \texttt{student, gpa, promoted} are only limited to constants of type \texttt{student}. Similarly, predicates \texttt{class, difficulty} are exclusive to the constants \texttt{english} and \texttt{math}. Type checking ensures that we do not consider ground atoms like \texttt{student(english)} or \texttt{class(phil)}.

\noindent Likewise for binary predicate \texttt{takes(S,C)}, the first variable is always grounded with a \texttt{student} type constant, and the second with a \texttt{class} type constant. Even in this miniature example, type checking reduces the number of ground atoms under consideration from 25 to only 6 - a 76\% reduction. Such gains significantly reduce complexity as size and sparsity of the graph increases.
\end{example}

\noindent\textbf{Detecting and Resolving Inconsistencies.} Inconsistency can occur in the following cases:
\begin{enumerate}
    \item For some ground atom, a new interpretation is assigned an annotation $[L',U']$ that is not a subset of the current interpretation $[L,U]$ (we assume $L \leq U$). i.e. if either $U < L'$ or $U' < L$.
    \item When an inconsistency occurs between an atom and its negation like ``a'' and ``not a''. Or between complementary predicates like ``$bachelor(X)$'' and ``$married(X)$'' which cannot hold simultaneously. \\
    e.g. Literal A has annotation $[L_1,U_1]$ and Literal B is the negation of literal A with annotation $[L_2,U_2]$. The fixpoint operator attempts to assign $[L_1',U_1']$ to Literal A, and $[L_2',U_2']$ to Literal B. But new bounds are inconsistent, i.e. either $L_1' > 1-L_2'$ or $U_1' < 1-U_2'$.
\end{enumerate}
PyReason flags all such inconsistencies arising during the execution of the fixpoint operator and reports them.  Further, as the fixpoint operator provides an explainable trace, the user can see the precise cause of the inconsistency.  As an additional, practical feature, PyReason includes an option to reset the annotation to $[0,1]$ for any identified inconsistency and set the atom to static for the remainder of the inference process.  In this way, such inconsistencies cannot propagate further.  These initial capabilities provide a solid foundation for more sophisticated consistency management techniques such as providing for local consistency or iterative relaxation of the initial logic program.

\begin{example}[Detecting and Resolving Inconsistencies.] Consider we have the following prior knowledge:

\texttt{I(takes(phil,math), 4) = [1,1]}

\texttt{I(takes(mary,math), 4) = [1,1]}

\texttt{I(friend(phil,mary), 5) = [0,0]}

\noindent However, the following logical rule with grounding $S \gets phil$, $S' \gets mary$, $C \gets math$:

$friend(S,S'):[1,1] \gets_1 takes(S,C):[1,1] \wedge takes(S',C):[1,1]$ gets fired at $t=4$.

\noindent resulting in:

\texttt{I(friend(phil,mary), 5) = [1,1]}

\noindent But clearly this is an inconsistency as \texttt{I(friend(phil,mary), 5)} cannot be both $[0,0]$ and $[1,1]$ simultaneously. So, we conclude that at least one of those two interpretations must be incorrect. If there is no way to ascertain which is correct, we may resolve this logical inconsistency by setting:

\texttt{I(friend(phil,mary), s) = [0,1]} at $t = 5$.
\end{example}

\section{Implementation}
\label{sec:software_implementation}

We have endeavored to create a modern Python-based framework to support scalable yet correct reasoning.  We allow graphical input via convenient \textit{Graphml} format, which is commonly used in knowledge graph architectures. The python library Networkx is used to load and interact with the graph data. We are currently in the process of directly supporting Neo4j. The initial conditions and rules are entered in YAML format and we use memory-efficient implementation techniques to correctly capture semantic structures. We use the Numba open-source JIT compiler to translate many key operations into fast, optimized machine code while allowing the user to interact with Python and the aforementioned front-ends. Our implementation can support CPU parallelism, as evidenced by our experiments run on multi-CPU machines.

Our software stores interpretations in a nested dictionary. For computational efficiency and ease of use, our software allows specification of a range of time-points $T = {t_1, t_2, \ldots}$ instead of a single time-point $t$, for which an interpretation $I$ remains valid. To reduce memory requirements, only the one set of interpretations (current) are stored at any point in time. However, past interpretations can be obtained using \textit{rule traces}, which retains the change history for each interpretation and the corresponding grounded logical rules that caused each change. \textit{Rule traces} make our software completely explainable, as every inference can be traced back to the cascade of rules that led to it.

MANCaLog~\cite{mancalog13} showed the use of the fixpoint operator for both canonical and non-canonical models. By recomputing interpretations at every time step, we not only require significantly less memory but also, support both the canonical and the non-canonical cases. Due to this design, increase in computation time is observed to be minimal.

Furthermore, we make significant advances on~\cite{PAREDES2021232} by supporting static predicates, and having in-built capabilities for non-graph reasoning, and type checking as detailed in section~\ref{sec:logical_framework}.

Our implementation can be found online as specified in section~\ref{sec:online_resources} and detailed pseudo-code can be found in the supplemental information.

\section{Experiments}
\label{sec:experimental_results}

\subsection{Honda Buyer-Supplier Dataset}
\label{sec:honda-expt}
We conduct our experiment on a Honda Buyer-Supplier network~\cite{yan_theory_2015}.  The dataset (network) contains 10,893 companies (nodes) and 47,247 buyer-supplier relationships between them (edges).

We design an use case, where we assume that operations of all companies from a particular country are disrupted, and observe the effects that this may have on companies across the world. We feel this is akin to supply chain issues faced worldwide during the COVID-19 pandemic. For our tests, we use the following logical rule which in practice would be either learned or come from an expert.\\
\begin{small}
    $disrupted(Buyer):[1,1] \gets_{\Delta t=1} \forall_{k} supplies(Sup_k, Buyer):[1,1], \exists_{k/2} disrupted(Sup_k):[1,1]$
\end{small}\\
It states that, a company is disrupted at a particular timestep if at least 50\% of its suppliers are totally disrupted in the previous timestep. We conduct this experiment for three different countries (USA, Taiwan, and Australia), having a wide range of proportion of companies in the dataset. We do not fix the number of inference steps, instead we let the diffusion process run until it converges (in bold). The results are shown in Table \ref{tab:honda_usecase}.
 
\begin{table*}
\caption{Honda network: How disruption on a country's industry, caused by a pandemic, may spread worldwide}
\label{tab:honda_usecase}
\begin{tabular}{lccccccccccc}
\toprule
\multicolumn{2}{c}{Companies} & \multicolumn{7}{c}{Companies disrupted across the world at time t=} & \multicolumn{3}{c}{\% of companies disrupted}\\
\cmidrule(lr){1-2} \cmidrule(lr){3-9} \cmidrule(lr){10-12}
Based & Count & 0 & 1 & 2 & 3 & 4 & \ldots & 38 & Initial & Final & \textbf{Change}\\
\midrule
USA & 1599 & 1599 & 1965 & 2057 & 2203 & 2313  & \ldots & \textbf{3336} & 14.68 & 30.75 & \textbf{16.07} \\
Taiwan & 603 & 603 & 644 & \textbf{647} & 647 & 647  & \ldots & 647 & 5.54 & 5.94 & \textbf{0.40} \\
Australia & 128 & 128 & \textbf{131} & 131 & 131 & 131  & \ldots & 131 & 1.18 & 1.21 & \textbf{0.03} \\
\bottomrule
\end{tabular}
\end{table*}

To test if our approach could scale, we use two inference rules which jointly state, a company is disrupted at a particular timestep if any of its supplier(s) are completely disrupted in the previous timestep, or if at least 50\% of its suppliers are disrupted to at least 50\% of their capacity. We conduct this experiment for different graph sizes, and for different number of timesteps to show the scaling capability of our software in Table \ref{tab:scalabilityExpt}.

\begin{table}
\caption{Scalability of our framework}
\label{tab:scalabilityExpt}
\begin{tabular}{ccccccc}
\toprule
Nodes (N) & Edges (E) & Total attributes & Density                           & Timesteps & Runtime (in s) & Memory (in MB) \\ \midrule
1000      & 410       & 5012             & 4.10 x $10^{-4}$                  & 2         & 0.36           & 4.9          \\
          &           &                  &                                   & 5         & 0.42           & 1.8          \\
          &           &                  &                                   & 15        & 0.34           & 0.1          \\ \midrule
2000      & 1640      & 13269            & 4.10 x $10^{-4}$                  & 2         & 0.43           & 1.2          \\
          &           &                  &                                   & 5         & 0.55           & 2.1          \\
          &           &                  &                                   & 15        & 0.81           & 8.2          \\ \midrule
5000      & 10244     & 57852            & 4.10 x $10^{-4}$                  & 2         & 1.54           & 17.2          \\
          &           &                  &                                   & 5         & 1.84           & 16.0         \\
          &           &                  &                                   & 15        & 3.38           & 54.6          \\ \midrule
10000     & 41034     & 197752           & 4.10 x $10^{-4}$                  & 2         & 4.83          & 80.3          \\
          &           &                  &                                   & 5         & 6.29          & 60.3
          \\
          &           &                  &                                   & 15        & 12.34          & 210.8
          \\
\bottomrule
\end{tabular}
\end{table}

The results show that both runtime and memory remain almost constant over large ranges, and then scale sub-linearly with increase in network size.

\subsection{Pokec Social Media dataset}
\label{sec:pokec-expt}
Pokec is a popular slovakian social network, and this dataset~\cite{takac2012data} contains personal information like gender, age, pets (attributes) of 1.6 million people (nodes), and 30.6 million connections between them (edges).

We take inspiration from the advertising community to design our use case. We consider, a small proportion of the population, who has pet(s), to be customers of a pet food company. The company, using Pokec data, must identify relevant advertising targets among the population. A realistic strategy can be captured by two logical rules:
\begin{enumerate}
    \item 
    \begin{small}
        $\forall X, Y relevance(X):[0.6,1] \gets_{\Delta t=1} relevance(Y):[1,1] \wedge friend(X, Y):[1,1]$
    \end{small}

    Friend of a relevant target or existing customer (always relevant), is at least 60\% relevant.

    \item
    \begin{small}
        $\forall X, Y relevance(X):[1,1] \gets_{\Delta t=1} relevance(Y):[1,1] \wedge friend(X, Y):[1,1] \wedge hasPet(X, P):[1,1] \wedge hasPet(Y, P):[1,1] $
    \end{small}

    Friend of a relevant target is totally relevant if they have pet(s) of same kind - dog, cat, \ldots

\end{enumerate}

The diffusion process converged after 8 timesteps, took 42 minutes to complete and used 58.36 GB of memory - which further showcases the scalability of our framework. The results are shown in Table \ref{tab:pokec_usecase}.
 
\begin{table}
\caption{Pokec social media: How brands may use consumer data to identify prospective customers}
\label{tab:pokec_usecase}
\begin{tabular}{ccccc}
\toprule
&&&\multicolumn{2}{c}{Advertising targets} \\
\cmidrule(lr){4-5}
Population size & Current Customers & Timesteps & Fully relevant & Partially relevant \\
\midrule
1,632,803 & 2,308 & 0 & 2,308 & 0 \\
&& 1 & 2,596 & 39,836 \\
&& 2 & 2,657 & 47,405 \\
&& 3 & 2,679 & 49,174 \\
&& 4 & 2,690 & 50,046 \\
&& 5 & 2,692 & 50,412 \\
&& 6 & 2,693 & 50,455 \\
&& 7, 8, \ldots & 2,693 & 50,608 \\
\bottomrule
\end{tabular}
\end{table}

The process of inference is completely explainable, and an user may use \textit{rule traces}, an optional output of PyReason, to identify the logical rules that led to change in each interpretation. An example of a rule trace from the previous experiment is presented in Table~\ref{tab:short_rule_trace}.

\begin{table*}
\caption{Rule trace for a single node for label \texttt{relevance}. Application of rule 1 above caused the first change from $[0,1]$ to $[0.6,1]$, followed by, an update to $[1,1]$ due to firing of rule 2. A list of node and edge IDs which were used to ground the rule clauses are also provided.}
\label{tab:short_rule_trace}
\begin{tabular}{p{0.01\textwidth}p{0.06\textwidth}p{0.06\textwidth}p{0.05\textwidth}cccc}
\toprule
t & Old Bound & New Bound & Rule fired & Clause-1 & Clause-2 & Clause-3 & Clause-4 \\
\midrule
1 & {[}0.0,1.0{]} & {[}0.6,1.0{]} & rule\_1 & {[}`354455'{]} & {[}(`354365', `354455'){]} &  &  \\
2 & {[}0.6,1.0{]} & {[}1.0,1.0{]} & rule\_2 & {[}`354455', & {[}(`354365', `354455'), & {[}(`718503', & {[}(`354365', \\
& & & &  `718503'{]} &  (`354365', `718503'){]} &  `cat'){]} &  `cat'){]}\\
\bottomrule
\end{tabular}
\end{table*}

All experiments were performed on an AWS EC2 container with 96 vCPUs (48 cores) and 384GB memory.

\section{Related work}

In section~\ref{sec:introduction}, we discussed how PyReason extends on the early modern logic programming languages like Prolog~\cite{colmerauer1982prolog}, Epilog~\cite{schubert2000episodic} and Datalog~\cite{abiteboul1995foundations} by supporting annotations. Recent neuro symbolic frameworks show great promise in the ability to learn or modify logic programs to align with historical data and improve robustness to noise.  Many such frameworks rely on an underlying differentiable, fuzzy, first order logic.  For example, logical tensor networks~\cite{ltn22} uses differentiable versions of fuzzy operators to combine ground and non-ground atomic propositions while logical neural networks~\cite{lnn2020} associate intervals of reals with atomic propositions and uses special parameterized operators.  Meanwhile, induction approaches such as differentiable ILP~\cite{deepMinIlp2018} fuzzy logic programs (using the product t-norm)  are learned from data based on template rule structures in a manner that support recursion and multi-step inference. In \cite{lnnInduction22}, Logical Neural Networks was used interpret learned rules in a precise manner. Here also, gradient descent was used to train the parameters of the network.
In the last two years, two paradigms have emerged with much popularity in the neuro symbolic literature.  Logical Tensor Networks (LTN)~\cite{ltn22} extend neural architectures through fuzzy, real-valued logic.  Logical Neural Networks (LNN)~\cite{lnn2020} provide a neuro symbolic framework with parameterized operators that supports open world reasoning in the logic.  As stated earlier, both can be viewed as a subset of annotated logic.  Hence, PyReason can be used to conduct inference on the logic for both frameworks, in addition to providing key capabilities such as graph-based and temporal reasoning, which currently are not present in the logics of those frameworks.

In both the forward pass of various neuro symbolic frameworks~\cite{neurASP2020, lnn2020, ltn22}, as well as for subsequent problems (e.g., entailment, abductive inference, planning, etc.), a deduction process is required. PyReason is designed to provide this precise capability.  Generalized annotated programs~\cite{ks92} has been shown to capture a wide variety of real-valued, temporal, and fuzzy logics as it associates logical atoms with elements of a lattice structure as opposed to scalar values.  As a result it can capture all the aforementioned logics, while retaining polynomial-time deduction due to the monotonicity of the lattice.  The use of a lattice structure allows for us to associate logical constructs with intervals, thus enabling open world reasoning. In our recent work~\cite{ssTAI22}, we provided extensions to~\cite{ks92} that allows for a lower lattice structure for annotations.  This enables the framework to capture paradigms such as LNN~\cite{lnn2020} and the MANCALog~\cite{mancalog13} for graph-based reasoning.  However, that work only showed that analogs to the theorems of~\cite{ssTAI22} for the lower lattice case and did not provide an implementation or experimental results.

By supporting generalized annotated logic, and its various extensions PyReason enables system design that is independent of the learning process.  As a result, once a neuro symbolic learning process creates or modifies a logic program based on data, PyReason can be used to efficiently answer deductive queries (to include entailment and consistency queries) as well as support more sophisticated inference such as abductive inference or planning.

Today knowledge graphs are crucial in representing data for reasoning and analysis. Recent research on creation of knowledge graphs~\cite{smajevic2021conceptual, glaser2022model} proposes methods to automatically convert conceptual models into knowledge graphs in GraphML format for enterprise architecture and a wide range of applications. PyReason, which supports the graphml format, could be an effective tool to reason about knowledge graphs obtained from one of these platforms.

\section{Conclusion and Future Work}

In this paper, we presented PyReason: an explainable inference software supporting annotated, open world, real-valued, graph-based, and temporal logics. Our modern implementation extends established generalized annotated logic framework to support scalable and efficient reasoning over large knowledge graphs and diffusion models. We are currently working on a range of extensions to this work. This includes adding more temporal logic operators for specification checking, learning rules from data through induction, and using the inference process to create new knowledge in non-static graphs (e.g., adding nodes and edges). We will also look to explore how PyReason can be used in conjunction with LTN~\cite{ltn22}, and LNN~\cite{lnn2020}. In supporting frameworks such as these,  we will look to add capabilities for symbol grounding~\cite{harnad_symbol_1990}, leveraging the results of the training process from frameworks such as LTN. Finally, we also plan on extending PyReason to act as a simulator for reinforcement learning based agents.

\begin{acknowledgments}
The authors are supported by internal funding from the Fulton Schools of Engineering and portions of this work is supported by U.S. Army Small Business Technology Transfer Program Office or the Army Research Office under Contract No.W911NF-22-P-0066.
\end{acknowledgments}

\bibliography{network.bib}

\appendix

\input{aaai_si.tex}

\end{document}

%% file: aaai_si.tex
\section{Formal Syntax and Semantics.}
\label{S1_Appendix}

We now recapitulate the definition of Generalized Annotated Logic programs (from now on referred to as ``GAPs'', for short) from \cite{ks92} as well as the extensions that we include in our software.

\begin{figure}[!h]
    \begin{center}
        \includegraphics[width=.9\linewidth]{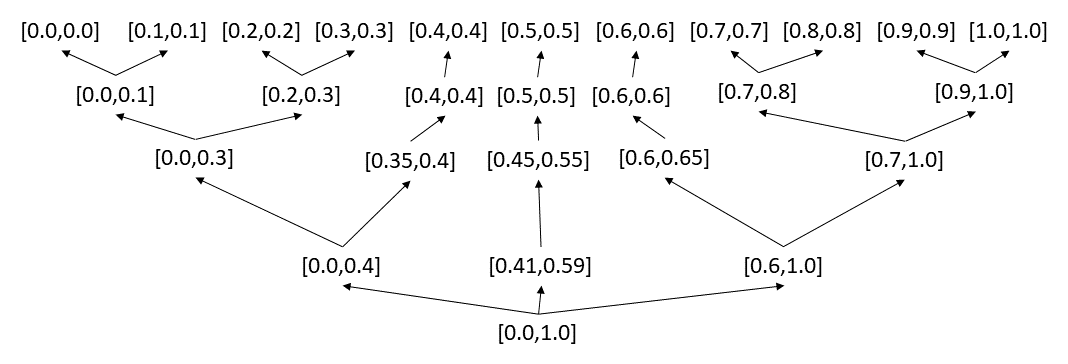}
    \end{center}
    \caption{\label{fig:lowerLattice}Example of a lower semilattice structure where the elements are intervals in $[0,1]$.}
\end{figure}

In \cite{ks92}, the authors assumed the existence of a semilattice. $\calt$ (not necessarily complete) with ordering $\sqsubseteq$.  To support contemporary applications in neuro symbolic reasoning \cite{lnn2020,deepMinIlp2018,ssTAI22, nttIlp21,lnnInduction22} as well as social network analysis \cite{mancalog13,tplp13} we implemented this as a lower semilattice structure. Therefore, we have a single element $\bot$ and multiple top elements $\top_0,\ldots\top_i\ldots\top_{max}$.  The notation $height(\calt)$ is the maximum number of elements in the lattice in a path between $\bot$ and a top element (including $\bot$ and the top element)\footnote{In general, we shall assume that the lattice consists of finite, discrete elements.}.  The employment of a lower semilattice structure allows enables two desirable characteristics.  First, we desire to annotate atoms with intervals of reals in $[0,1]$ as done in previous work~\cite{lnn2020,mancalog13,apt11}.  Second, it allows for reasoning about such intervals whereby the amount of uncertainty (i.e., for interval $[l,u]$ the quantity $u-\ell$) decreases monotonically as an operator proceeds up the lattice structure.  Therefore, we define bottom element $\bot = [0,1]$ and a set of top elements $\{[x,x] \; | \; [x,x]\subseteq [0,1]\}$ (see note\footnote{N.B. that when using a semilattice of bounds, the notation ``$\sqsubseteq$'' loses its ``subset intuition'', as $[0,1] \sqsubseteq [1,1]$ in this case, for example.}).  Specifically, we set $\top_0 = [0,0]$ and $\top_{max}=[1,1]$.  An example of such a semilattice structure is shown in Figure~\ref{fig:lowerLattice}.

As with~\cite{ks92}, we assume the existence of a set $\avar$ of variable symbols ranging over $\calt$ and a set $\calf$ of function symbols, each of which has an associated arity. We start by defining annotations.

\begin{definition}[Annotation]
\textsf{(i)} Any member of $\calt\,\cup \avar$ is an annotation.\\
\textsf{(ii)} If $f$ is an $n$-ary function symbol over $\calt$ and $t_1,\ldots,t_n$ are annotations, then $f(t_1,\ldots,t_n)$ is an annotation.
\end{definition}

One specific function we define is ``$\neg$'' which is used in semantics of \cite{ks92} as well as the more recent interval-based framework used in \cite{lnn2020}.  For a given $[l,u]$, $\neg([l,u])=[1-u, 1-l]$.  Note that we also use the symbol  $\neg$ in our first-order language (following the formalism of \cite{ks92}).  We define a separate logical language whose constants are members of set $\calc$ and whose predicate symbols are specified by set $\calp$.  We also assume the existence of a set $\calv$ of variable symbols ranging over the constants, that no function symbols are present, and terms and atoms are defined in the usual way (cf.\ \cite{ll87}).  We shall assume that $\calc,\calp,\calv$ are discrete and finite.  In general, we shall use capital letters for variable symbols and lowercase letters for constants.  Similar to previous work~\cite{deepMinIlp2018,thomasNsr21} shall assume that all elements of $\calp$ have an arity of either 1 or 2 - so we shall denote these $\calp_{una}$ for unary predicates and $\calp_{rel}$ for binary predicates.   We shall also denote a subsets of $\calp$ to include ``target predicates'' written $\calp_{tgt}$ that can consist of either binary or unary predicates ($\calp_{tgt\_rel},\calp_{tgt\_una}$) provided that they are not reserved words.  We shall use the symbol $\call$ to denote the set of all ground literals and $\cala$ for the set of all ground atoms.  We now define the syntactical structure of GAPs that will be used in this work.

\begin{definition}[Annotated atoms, negations, literals]
The core syntactic structures are defined as follows:
\begin{itemize}
\item{\textbf{Annotated atom.} If $a$ is an atom and $\mu$ is an annotation, then $a:\mu$ is an \emph{annotated atom}.}
\item{\textbf{Annotated Negation.} If $a$ is an atom and $\mu$ is an annotation, then $\neg a:\mu$ is an \emph{annotated negation}.}
\item{\textbf{Annotated Literal.} Collectively, atoms and negations are referred to as \emph{annotated literals}.}
\end{itemize}
\end{definition}

\begin{definition}[GAP Rule]
If $\ell_0:\mu_0, \ell_1:\mu_1,\ldots,\ell_m:\mu_m$ are annotated literals (such that for all $i,j \in 1,m$, $\ell_i\not\equiv \ell_j$), then
\begin{eqnarray*}
r\equiv \ell_0:\mu_0 & \leftarrow & \ell_1:\mu_1\,\wedge\,\ldots\wedge\, \ell_m:\mu_m
\end{eqnarray*}
is called a \emph{GAP rule}.  We will use the notation $head(r)$ and $body(r)$ to denote $\ell_0$ and $\{\ell_1,\ldots,\ell_m\}$ respectively. When $m=0$ ($body(r)=\emptyset$), the above GAP-rule is called a \emph{fact}.  A GAP-rule is \emph{ground} iff there are no occurrences of variables from either $\avar$ or $\calv$ in it.  For ground rule $r$ and ground literal $\ell$, $bodyAnno(\ell,r)= \mu$\textit{ such that }$\ell:\mu$\textit{ appears in the body} of $r$.  A generalized annotated program $\Pi$ is a finite set of GAP rules.
\end{definition}

The formal semantics of GAPs are defined as follows.  Note that we extend the notion of an interpretation to allow for a mapping of literals to annotations (as opposed to atoms).  However, we add a requirement on the annotation between each atom and negation that ensures equivalence to the semantic structure of \cite{ks92}.

\begin{definition}[Interpretation]
\label{def:interp}
An interpretation $I$ is any mapping from the set of all grounds literals to $\calt$ such that for literals $a, \neg a$, we have $I(a) = \neg(I(\neg a))$.  The set $\cali$ of all interpretations can be partially ordered via the ordering: $I_1\preceq I_2$ iff for all ground literals $a$, $I_1(\ell)\sqsubseteq I_2(\ell)$. $\cali$ forms a complete lattice under the $\preceq$ ordering.
\end{definition}

Now we present the satisfaction relationship:

\begin{definition}[Satisfaction]
An interpretation $I$ \emph{satisfies} a ground literal $\ell:\mu$, denoted $I\models \ell:\mu$, iff $\mu \sqsubseteq I(\ell)$. $I$ satisfies the ground GAP-rule

\begin{eqnarray*}
\ell_0: \mu_0 & \leftarrow & \ell_1:\mu_1\wedge\,\ldots\,\wedge\, \ell_m:\mu_m
\end{eqnarray*}

(denoted $I\models \ell_0:\mu_0\leftarrow \ell_1:\mu_1\,\,\wedge\,\ldots\,\wedge\,\ell_m: \mu_m$)  iff either

\begin{enumerate}
\item $I$ satisfies $\ell_0:\mu_0$ or
\item There exists an $1\leq i\leq m$ such that $I$ does not satisfy $\ell_i:\mu_i$.
\end{enumerate}

$I$ satisfies a non-ground literal or rule iff $I$ satisfies all ground instances of it.
\end{definition}

We say that an interpretation $I$ is a \emph{model} of program $\Pi$ if it satisfies all rules in $\Pi$.  
Likewise, program $\Pi$ is \emph{consistent} if there exists some $I$ that is a model of $\Pi$.  
We say $\Pi$ \emph{entails} $\ell:\mu$, denoted $\Pi\models \ell:\mu$, iff for every interpretation $I$ s.t.\ $I\models \Pi$, we have that $I\models \ell:\mu$.  
As shown by \cite{ks92}, we can associate a fixpoint operator with any GAP $\Pi$ that maps interpretations to interpretations.

\begin{definition}
Suppose $\Pi$ is any GAP and $I$ an interpretation.
The mapping $\T_{\Pi}$ that maps interpretations to interpretations is defined as
\[
\T_{\Pi}(I)(\ell_0) = \mathbf{sup}(annoSet_{\Pi,I}(\ell_0)),
\]
where $annoSet_{\Pi,I}(\ell_0)= \{I(\ell_0) \}\cup\{ \mu_0\: \; | \; \: \ell_0:\mu_0\leftarrow \ell_1:\mu_1\,\wedge\ldots\wedge\, \ell_m:\mu_m \textit{ is a ground}$ $\textit{instance of a rule in } \Pi,
\textit{ and for all } 1\leq  i\leq m, \textit{we have } I\models \ell_i:\mu_i\}$
\end{definition}

The key result of \cite{ks92} tells us that $\mathit{lfp}(\T_{\Pi})$ precisely captures the ground atomic logical consequences of $\Pi$.  We show this is also true (under the condition that $\Pi$ is consistent) even if the annotations are based on a lower lattice.  In \cite{ks92}, the authors also define the \emph{iteration} of $\T_\Pi$ as follows:

\begin{itemize}
\item $\T_\Pi\uparrow 0$ is the interpretation that assigns $\bot$ to all ground literals.
\item $\T_\Pi\uparrow (i+1) = \T_\Pi(\T_\Pi\uparrow i)$.
\end{itemize}

For each ground $\ell \in \call$, the set $\Pi(\ell)$ is the subset of ground rules (to include facts) in $\Pi$ where $\ell$ is in the head.  We will use the notation $m_\ell$ to denote the number of rules in $\Pi(\ell)$.  
For a given ground rule, we will use the symbol $r_{\ell,i}$ to denote that it is the $i$-th rule with atom $\ell$ in the head.

\section{Formal Proofs for Results where the Lower Lattice Assumption is Made.}
\label{S2_Proofs}

Please see \cite{ssTAI22}.

\section{Additional Details on Supply Chain Experiments}
\label{S7_Appendix}

Table~\ref{tab:honda_data} summarizes the contents of the dataset. 7,396 companies were listed with a unique GICS, while 3,497 did not have one listed.

\begin{table}
\caption{Overview of the Honda Buyer-Supplier Network}
\label{tab:honda_data}
\begin{tabular}{lc} 
\toprule
Companies (Nodes) & 10,893\\ 
Buey-Supplier Relationships (Edges) & 47,247\\
Global Industry Classification Standard (GICS) types & 67\\
Edge Relationship types & 4\\
\bottomrule
\end{tabular}
\end{table}

\begin{figure}[b]
\centering
\includegraphics[width=0.7\linewidth]{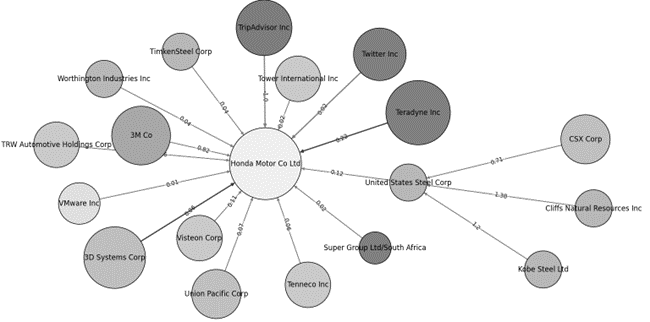}
\caption{A snapshot of the network showing link connections with attribute labels}
\label{fig:dataNet2}
\end{figure}

To understand the structure of the buyer-supplier network more in-depth, one industry along with its first and second Tier suppliers are mapped, as shown in Figure \ref{fig:dataNet2}. Each node shown in the figure contain an attribute \emph{name} which describes the name of the industry and each edge connecting those industries contain an attribute \emph{cost} which describes the profit relationship between those two particular industries connected by an edge.

\begin{figure}
\centering
\includegraphics[scale=0.45]{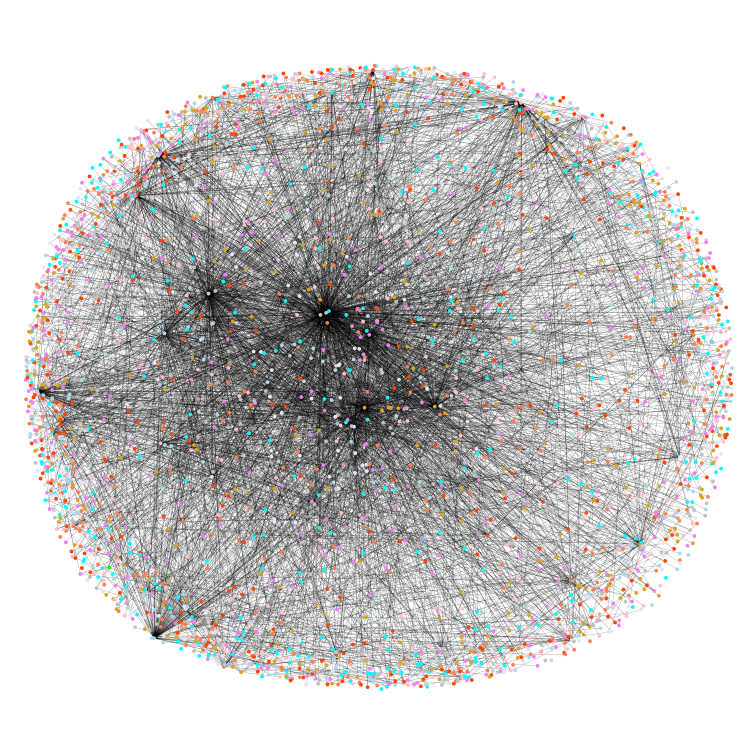}
\caption{Honda Supply Chain Network. Each node and edge is differently colored based on the type.}
\label{fig:dataNet1}
\end{figure}

In Figure~\ref{fig:dataNet1} a portion of data is mapped out showing the complexity of the supply network. The figure can help identify a few key nodes which connect large sections of the network to other major sections - which makes them critical to operations.

\begin{figure}
\centering
\includegraphics[scale=0.4]{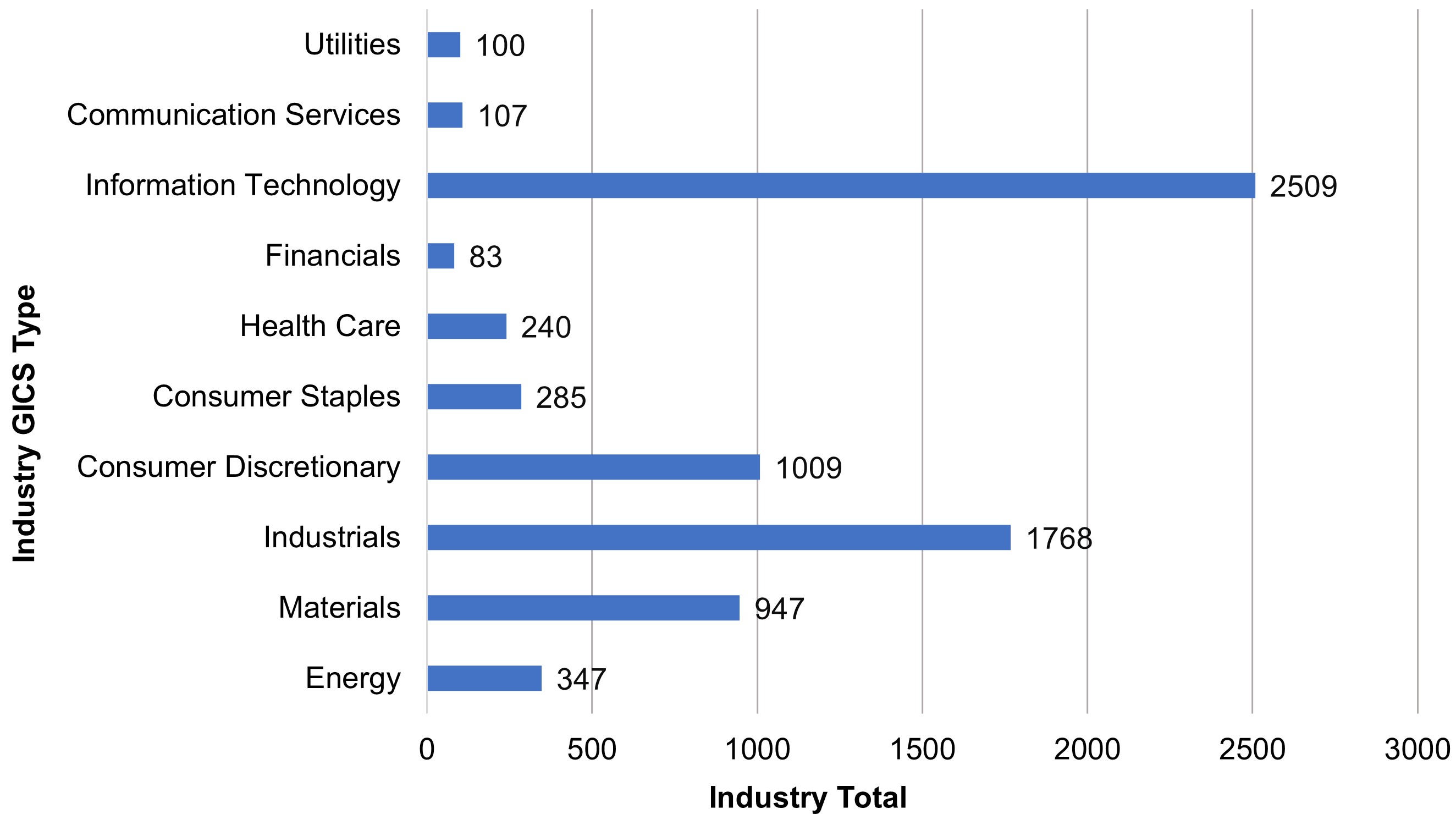}
\caption{Distribution of the nodes in the network based on its GICS type.}
\label{fig:gicsGraph}
\end{figure}

\begin{figure}
\centering
\includegraphics[scale=0.4]{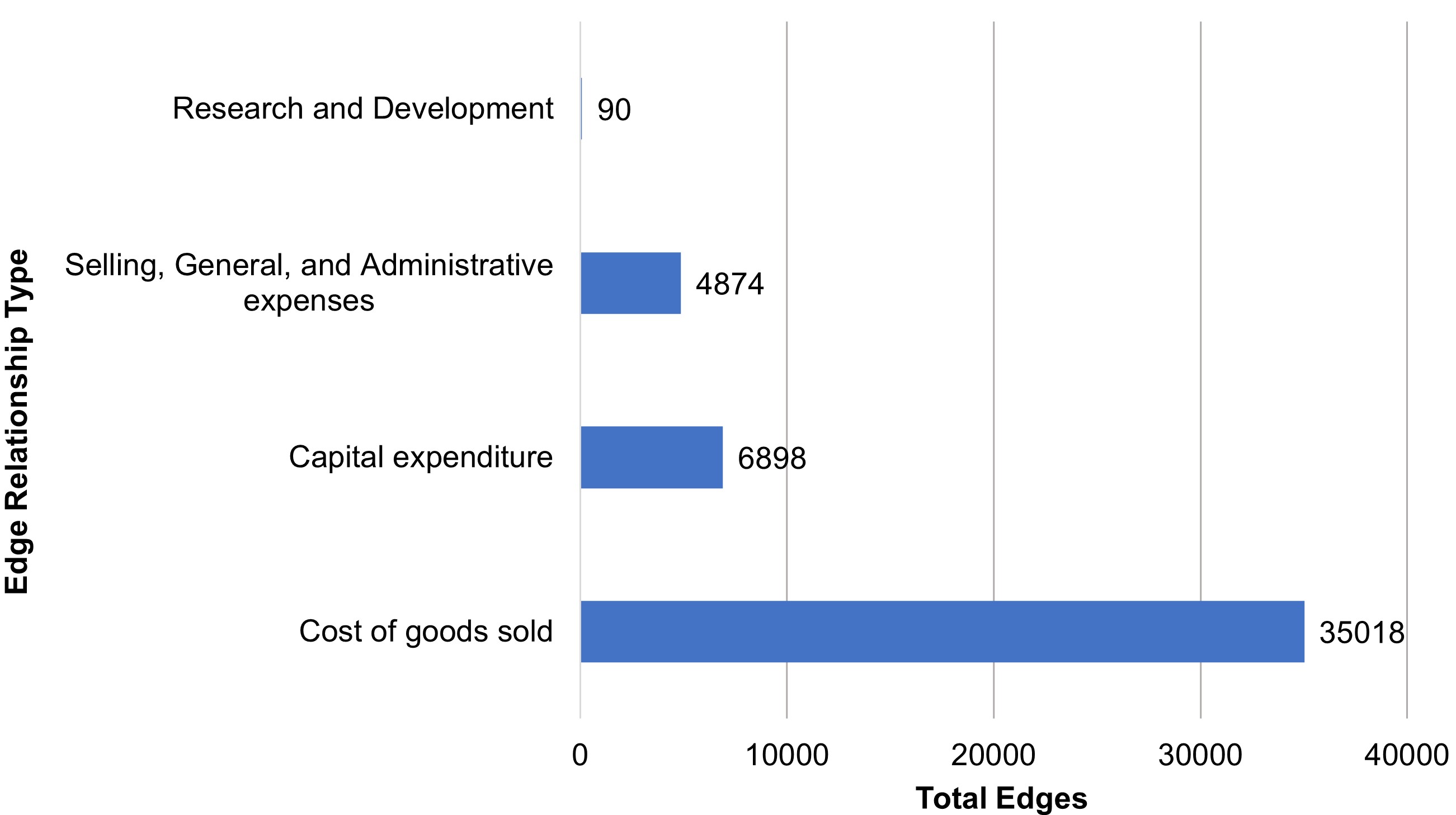}
\caption{Edge relationships in the dataset.}
\label{fig:edgesGraph}
\end{figure}

Figure~\ref{fig:gicsGraph} shows the distribution of various industry types. Figure~\ref{fig:edgesGraph} gives different edge relationships present in the dataset.

\section{Additional Details on Social Network Experiments}
\label{S8_Appendix}

\begin{figure}[b]
    \centering
    \includegraphics[width=0.8\linewidth]{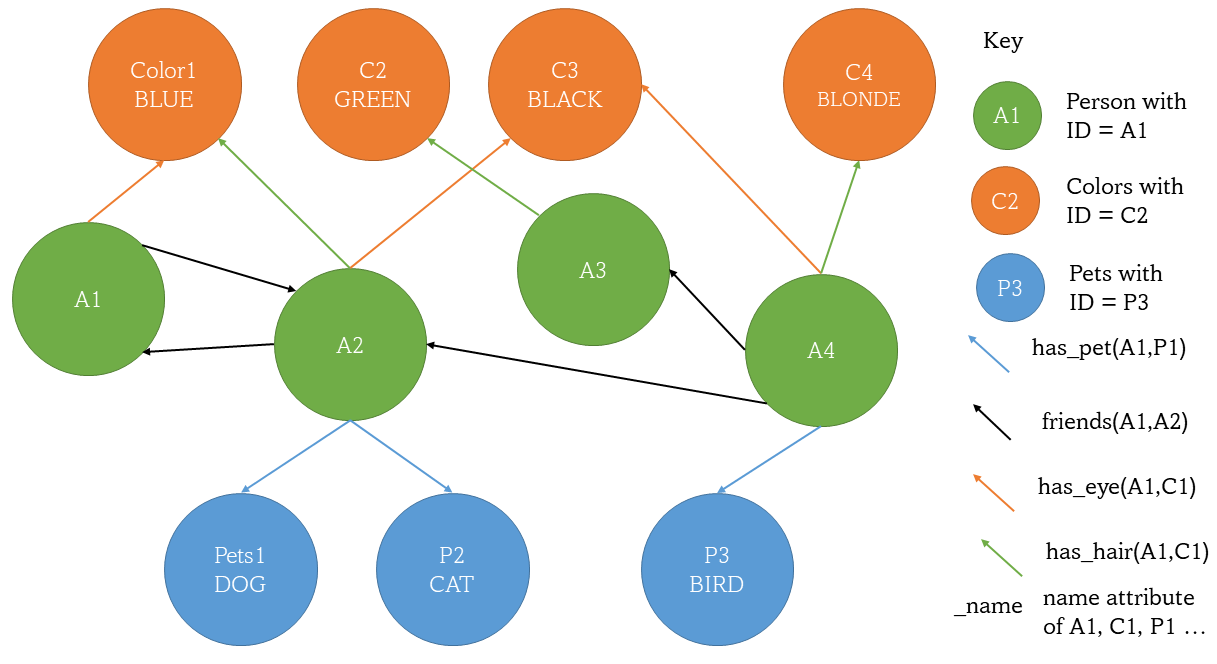}
    \caption{Schema of Pokec network}
    \label{fig:pokec_schema}
\end{figure}

Table~\ref{tab:pokec_data} gives an overview of the dataset, and the schema is shown in Fig.~\ref{fig:pokec_schema}. The edge relationship types are enumerated in the figure.

\begin{table}[t]
\caption{Overview of the Pokec Social Media dataset}
\label{tab:pokec_data}
\begin{tabular}{lc} 

\toprule
Nodes & 1,632,820\\ 

Edges & 31,633,113\\

Node attribute types & 5\\

Edge Relationship types & 4\\
\bottomrule
\end{tabular}
\end{table}

Graph density represents the ratio between the edges present in a graph and the maximum number of edges that the graph can contain. In reality, graphs are often sparse (not dense). To test the scaling capacity of our approach with graph density, we re-run the experiment in Table~\ref{tab:scalabilityExpt} while modifying the density of the dataset. The results, in Table~\ref{tab:sparsity_expt}, show that the experiments on Honda data, which is about 40 times more dense than the Pokec network takes only about 3 times more memory, and 6 times more runtime to complete. This further showcases the scaling capability of our framework.

\begin{table}
\caption{Impact of graph density on memory and runtime}
\label{tab:sparsity_expt}
\begin{tabular}{cccccc}
\toprule
& & \multicolumn{2}{c}{Honda} & \multicolumn{2}{c}{Pokec} \\
& Density$\rightarrow$ & \multicolumn{2}{c}{4.10 x $10^{-4}$} & \multicolumn{2}{c}{0.11 x $10^{-4}$} \\
 \cmidrule(lr){3-4} \cmidrule(lr){5-6}
Nodes (N) & Timesteps & Runtime (in s) & Memory (in MB) & Runtime (in s) & Memory (in MB) \\ \midrule

10000     & 2         & 4.83          & 80.3   & 0.70   & 27.4    \\
          & 5         & 6.29          & 60.3   & 0.93   & 19.9    \\
          & 15        & 12.34         & 210.8  & 1.73   & 21.2   \\
\bottomrule
\end{tabular}
\end{table}

Finally, to test the scalability of our approach with respect to the number of attributes (and hence, number of ground atoms) in a graph, we run an experiment for a simple use-case - once with the original dataset, and then with the attributes added in. We define the use-case on the social media data as:\\
$\forall X, Y ~~infected(X):[1,1] \gets_{\Delta t=1} infected(Y):[1,1] \wedge friend(X, Y):[1,1]$\\
which says, ``A person contracts the virus if any of their friend(s) is infected by a virus''. The results are as shown in Table~\ref{tab:attributes_expt}.

\begin{table}
\caption{Impact of adding 7 attributes on memory and runtime for 5 timesteps}
\label{tab:attributes_expt}
\begin{tabular}{lccccc}
\toprule
Attributes & Nodes (N) & Edges (E) & Grounded atoms & Runtime (in mins) & Memory (in GB)\\ \midrule
Not added & 1,632,803 & 30,622,563 & 3,265,606  & 19.95 & 42.44\\
Added & 1,632,820 & 31,633,113 & 36,531,539  & 28.26 & 59.39\\
Change & (+17) & (+1,010,550) & (+33,265,933)  & (+8.31) & (+16.95)\\
\bottomrule
\end{tabular}
\end{table}

\section{Expanded Rule Traces}
\label{S4_rule_traces}

A longer version of the rule trace in Table~\ref{tab:short_rule_trace}, with 10 atoms is presented in Table~\ref{tab:long_rule_trace}.

\begin{table*}
\caption{A longer version of the rule trace from Table~\ref{tab:short_rule_trace} for \texttt{Label:relevance}}
\label{tab:long_rule_trace}
\begin{tabular}{p{0.01\textwidth}p{0.01\textwidth}p{0.06\textwidth}p{0.06\textwidth}p{0.06\textwidth}p{0.05\textwidth}p{0.1\textwidth}p{0.22\textwidth}p{0.12\textwidth}p{0.12\textwidth}}
\toprule
&&&\multicolumn{2}{c}{Bound}&&&&& \\
\cmidrule(lr){4-5}

t & $\Gamma$ & Node & Old & New & Rule fired & Clause-1 & Clause-2 & Clause-3 & Clause-4 \\
\midrule
1 & 1 & 1273439 & {[}0.0,1.0{]} & {[}1.0,1.0{]} & rule\_2 & {[}`835886'{]} & {[}(`1273439', `835886'){]} & {[}(`835886', `cat'){]} & {[}(`1273439', `cat'){]} \\
1 & 1 & 103308 & {[}0.0,1.0{]} & {[}0.6,1.0{]} & rule\_1 & {[}`659792', `404372'{]} & {[}(`103308', `659792'), (`103308', `404372'){]} &  &  \\
\midrule
2 & 2 & 277684 & {[}0.0,1.0{]} & {[}1.0,1.0{]} & rule\_2 & {[}`305645'{]} & {[}(`277684', `305645'){]} & {[}(`305645', `fish'){]} & {[}(`277684', `fish'){]} \\
2 & 2 & 551249 & {[}0.0,1.0{]} & {[}0.6,1.0{]} & rule\_1 & {[}`377195'{]} & {[}(`551249', `377195'){]} &  &  \\
\midrule
3 & 3 & 861455 & {[}0.0,1.0{]} & {[}1.0,1.0{]} & rule\_2 & {[}`1450147'{]} & {[}(`861455', `1450147'){]} & {[}(`1450147', `spider'){]} & {[}(`861455', `spider'){]} \\
3 & 3 & 23197 & {[}0.0,1.0{]} & {[}0.6,1.0{]} & rule\_1 & {[}`25795'{]} & {[}(`23197', `25795'){]} &  &  \\
3 & 3 & 757646 & {[}0.0,1.0{]} & {[}0.6,1.0{]} & rule\_1 & {[}`423053'{]} & {[}(`757646', `423053'){]} &  &  \\
\midrule
4 & 4 & 86436 & {[}0.0,1.0{]} & {[}1.0,1.0{]} & rule\_2 & {[}`743812'{]} & {[}(`86436', `743812'){]} & {[}(`743812', `cat'){]} & {[}(`86436', `cat'){]} \\
4 & 4 & 40242 & {[}0.0,1.0{]} & {[}0.6,1.0{]} & rule\_1 & {[}`407809'{]} & {[}(`40242', `407809'){]} &  &  \\
4 & 4 & 757646 & {[}0.0,1.0{]} & {[}1.0,1.0{]} & rule\_2 & {[}`423053', `548848'{]} & {[}(`757646', `423053'), (`757646', `548848'){]} & {[}(`548848', `cat'){]} & {[}(`757646', `cat'){]} \\
\midrule
5 & 5 & 420093 & {[}0.0,1.0{]} & {[}0.6,1.0{]} & rule\_1 & {[}`275269', `472129'{]} & {[}(`420093', `275269'), (`420093', `472129'){]} &  &  \\
5 & 5 & 1334826 & {[}0.0,1.0{]} & {[}1.0,1.0{]} & rule\_2 & {[}`1486432'{]} & {[}(`1334826', `1486432'){]} & {[}(`1486432', `fish'){]} & {[}(`1334826', `fish'){]} \\
5 & 5 & 196947 & {[}0.0,1.0{]} & {[}0.6,1.0{]} & rule\_1 & {[}`212129'{]} & {[}(`196947', `212129'){]} &  &  \\
\midrule
6 & 6 & 348252 & {[}0.0,1.0{]} & {[}1.0,1.0{]} & rule\_2 & {[}`1123497'{]} & {[}(`348252', `1123497'){]} & {[}(`1123497', `cat'){]} & {[}(`348252', `cat'){]} \\
6 & 6 & 1144981 & {[}0.0,1.0{]} & {[}0.6,1.0{]} & rule\_1 & {[}`232110'{]} & {[}(`1144981', `232110'){]} &  &  \\
6 & 6 & 420093 & {[}0.0,1.0{]} & {[}1.0,1.0{]} & rule\_2 & {[}`275269', `472129', `275337'{]} & {[}(`420093', `275269'), (`420093', `472129'), (`420093', `275337'){]} & {[}(`275337', `turtle'){]} & {[}(`420093', `turtle'){]} \\
\midrule
7 & 7 & 354365 & {[}0.0,1.0{]} & {[}1.0,1.0{]} & rule\_2 & {[}`354455', `718503'{]} & {[}(`354365', `354455'), (`354365', `718503'){]} & {[}(`718503', `cat'){]} & {[}(`354365', `cat'){]} \\
7 & 7 & 420093 & {[}0.0,1.0{]} & {[}1.0,1.0{]} & rule\_2 & {[}`275269', `472129', `275337'{]} & {[}(`420093', `275269'), (`420093', `472129'), (`420093', `275337'){]} & {[}(`275337', `turtle'){]} & {[}(`420093', `turtle'){]} \\
7 & 7 & 757646 & {[}0.0,1.0{]} & {[}1.0,1.0{]} & rule\_2 & {[}`423053', `548848'{]} & {[}(`757646', `423053'), (`757646', `548848'){]} & {[}(`548848', `cat'){]} & {[}(`757646', `cat'){]} \\
7 & 7 & 50219 & {[}0.0,1.0{]} & {[}1.0,1.0{]} & rule\_2 & {[}`2067', `50136'{]} & {[}(`50219', `2067'), (`50219', `50136'){]} & {[}(`2067', `cat'), (`50136', `cat'){]} & {[}(`50219', `cat'), (`50219', `cat'){]} \\
7 & 7 & 148995 & {[}0.0,1.0{]} & {[}0.6,1.0{]} & rule\_1 & {[}`140490'{]} & {[}(`148995', `140490'){]} &  &  \\
\bottomrule
\end{tabular}
\end{table*}

\section{Implementation Pseudocode}
\label{S5_algorithms}
Algorithm~\ref{alg:datastruc} enumerates the data structures in use. Algorithm~\ref{alg:init} shows the initial state, while algorithm~\ref{alg:flow} details the inference process. During inference, interpretations are updated as shown in algorithm~\ref{alg:update}. Logical consistency is maintained using algorithms~\ref{alg:conchk} and~\ref{alg:inconres}.

\begin{algorithm}
\caption{Data Structures Used}\label{alg:datastruc}
\begin{algorithmic}[1]
\State Nested Dictionary $\bm{I}=[Node/Edge,[Predicate, [Lower, Upper, Static]]]$ to store current interpretations only. If $Static$ is set to $1$, bounds:$Lower,Upper$ can no longer change for rest of program.
\State List $\bm{L}=[(Node/Edge, Predicate, Lower, Upper, Static,at\_t)]$ to store facts and inferences, before it is used to update the dictionary.

\State List $\textbf{IPL}=[(Predicate_1, Predicate_2)]$ containing pairs of predicates which cannot hold simultaneously, i.e., the bounds must be pairwise complementary. In the propositional case, if one of the predicates is $true$, the other must be $false$. We call this  ``inconsistent predicate list (IPL)''.
\State List $\bm{E}=[(Node/Edge, Predicate)]$ containing list of predicates that becomes inconsistent in the course of program execution.
\end{algorithmic}
\end{algorithm}

\begin{algorithm}
\caption{Program initialization}\label{alg:init}
\begin{algorithmic}[1]

\State $\bm{I}$ as follows:
\Statex $\forall$ nodes/edges, use $type\_checking$ to initialize valid predicates only.
\Statex All bounds are initialized to [0,1]. $Static$ set to 0.
\State $\bm{L} \gets [~]$ \Comment{Empty list}
\Statex Facts (incl. initial interpretations) are then copied into $\bm{L}$
\State $t \gets 0$
\State $\bm{E} \gets [~]$
\State Input: Number of diffusion time-steps $T$, Set of rules $\bm{R}$

\end{algorithmic}
\end{algorithm}

\begin{algorithm}
\caption{Program flow}\label{alg:flow}
\begin{algorithmic}[1]

\While{$t \leq T$}

\For{$i$ in $I$, where ($Static$ is $false$)}
    \State reset bounds to [0,1]
\EndFor

\State $update\_req \gets 0$
\For{$l$ in $\bm{L}$, where ($l(at\_t) == t$)}
    \If{check\_consistency($l \in \bm{L}$,$l \in  \bm{I}$)}
        \State update\_req += update\_interp($l \in \bm{L}$,$l \in  \bm{I}$)
    \Else
        \State resolve\_inconsistency($l \in  \bm{I}$)
        \If{$(l, l') \in \textbf{IPL}$, $\forall l'$}
            \State resolve\_inconsistency($l' \in  \bm{I}$)
        \EndIf
    \EndIf
\EndFor

\If{$update\_req$}
    \State Apply fix-point operator($gamma$) once.
    \For{each resulting interpretation}
        \If{$Static$ is $false$ in $I$}
            \State Add to $\bm{L}$
        \EndIf
    \EndFor
    \State Go to line $5$.
\Else
    \State $t \gets t+1$.
\EndIf

\EndWhile
\end{algorithmic}
\end{algorithm}

\begin{algorithm}
\caption{Updating interpretations}\label{alg:update}
\begin{algorithmic}[1]

\Procedure{update\_interp}{$i', i$}
     \State $updated \gets 0$
     \If{$i(Lower) != i'(Lower)~\text(or)~i(Upper) != i'(Upper)$}
        \State $i(Lower) \gets f_l(i(Lower), i'(Lower))$
        \Statex \Comment{by default $f_l$ is the $max()$ function, but it can be user defined.}
        \State $i(Upper) \gets f_u(i(Upper), i'(Upper))$
        \Statex \Comment{by default $f_u$ is the $min()$ function, but it can be user defined.}
        \State $updated \gets 1$
    \EndIf
    \If{$updated~\text(and)~(i, ic) \in \textbf{IPL}$, $\forall ic$}
        \State $ic(Lower) \gets f_l(ic(Lower), 1 - i(Upper))$
        \State $ic(Upper) \gets f_u(ic(Upper), 1 - i(Lower))$
    \EndIf
    \State return $updated$
\EndProcedure

\end{algorithmic}
\end{algorithm}

\begin{algorithm}
\caption{Consistency checking}\label{alg:conchk}
\begin{algorithmic}[1]

\Procedure{check\_consistency}{$i',i$}
\Statex \Comment{$i'$ is new interpretation with $[L', U']$, and, $i$ is current interpretation with $[L, U]$}
  \If{$L'>U~\text(or)~U'<L$}
    \State return $False$
  \Else
    \State return $True$
  \EndIf
\EndProcedure

\end{algorithmic}
\end{algorithm}

\begin{algorithm}
\caption{Inconsistency resolution}\label{alg:inconres}
\begin{algorithmic}[1]

\Procedure{resolve\_inconsistency}{$i \in \bm{I}$}
  \State $i(Lower) \gets 0$
  \State $i(Upper) \gets 1$
  \State $i(Static) \gets 1$
\EndProcedure

\end{algorithmic}
\end{algorithm}